\documentclass[aps,prl,twocolumn,superscriptaddress,showpacs]{revtex4}
\usepackage{amsfonts,amssymb,amsmath,latexsym,epsfig,times} 

\begin{document}

[Phys. Rev. Lett. {\bf 91}, 231101 (2003)]

\title{Relativistic chaos is coordinate invariant}

\author{Adilson E. Motter}
\email{motter@mpipks-dresden.mpg.de}
\affiliation{Max Planck Institute for the Physics of Complex Systems, N\"othnitzer Strasse 38,
01187 Dresden, Germany}

\date{\today}

\begin{abstract}
The noninvariance of Lyapunov exponents in general relativity has led
to the conclusion that chaos depends on the choice of the space-{\it
time} coordinates.  Strikingly, we uncover the transformation
laws of Lyapunov exponents under general space-time transformations
and we find that {\it chaos}, as characterized by positive Lyapunov
exponents, {\it is coordinate invariant}. As a result, the previous
conclusion regarding the noninvariance of chaos in cosmology, a major
claim about chaos in general relativity, necessarily involves the
violation of hypotheses required for a proper definition of the
Lyapunov exponents.
\end{abstract}

\maketitle

Chaotic properties of dynamical systems with a reparametrizable time
coordinate are important in physical theories without an absolute
time, such as general relativity. The study of chaos in general
relativity has followed two main lines. One considers the geodesic
motion of test particles in a given gravitational field \cite{bc92_vl96}.  The other
investigates the time evolution of the gravitational field itself \cite{b82,book:94},
which is relevant in cosmology. While the former case has been studied
with standard methods of the dynamical systems theory, an adequate
characterization of chaos in the latter is currently an open problem.
The difficulty comes from the dependence of the {\it shear} between
nearby trajectories on their time parametrization.  Accordingly,
dynamical properties, such as mixing and initial-condition
sensitivity, may depend on the time parametrization.

In classical physics, the study of dynamical systems
concerns differential equations of the form
\begin{equation} 
d\mathbf{ x}/dt=\mathbf{ F}(\mathbf{ x}),
\label{1}
\end{equation}
where $t$ is a uniquely defined parameter that usually represents the
time. Although a general definition of chaos is missing, it is widely
accepted that it regards the dynamics of bounded orbits and that a
chaotic system must present sensitive dependence on initial
conditions \cite{o94}.  Chaos can then be quantified in terms of Lyapunov
exponents \cite{o68} insofar as the following conditions are satisfied: ({\it i})
the system is {\it autonomous}; ({\it ii}) the relevant part of the
phase space is {\it bounded}; ({\it iii}) the invariant measure is
{\it normalizable}; ({\it iv}) the domain of the time parameter is
{\it infinite}.  Such a characterization is convenient because it is
invariant under space diffeomorphisms of the form $\mathbf{
y}=\boldsymbol{ \psi}(\mathbf{ x})$. As a result, chaos is a property
of the physical system and does not depend on the coordinates used to
describe the system.

In general relativity, the nonexistence of an absolute time parameter
forces us to consider Eq.~(\ref{1}) under space-{\it time}
diffeomorphisms: $\mathbf{ y}=\boldsymbol{ \psi}(\mathbf{ x},t)$,
$d\tau=\lambda(\mathbf{ x},t)dt$. A conceptual problem then arises
because of the dependence of classical indicators of chaos, like
Lyapunov exponents and entropies, on the choice of the time parameter.
This problem has attracted a great deal of attention since it was first
identified in the mixmaster cosmological model \cite{m69}, whose
largest Lyapunov exponent was shown to be positive or zero for
different choices of the coordinates \cite{fm88}. In
particular, numerous methods based on invariant curvature \cite{s93}, symbolic
dynamics \cite{r94}, Painlev\'e analysis \cite{cgr94}, and fractals \cite{bl98,ml01}, have
been proposed toward an invariant characterization of chaos in
cosmology.  The problem, however, goes beyond relativistic cosmology
since it has been argued that the same kind of noninvariance can be
observed in a system as simple as the harmonic oscillator if time
reparametrizations are allowed \cite{rb01}. Moreover, it has been exhibited
examples of systems whose nonmixing dynamics can be converted into a
mixing one through a time reparametrization \cite{f02}.  These
results have led to the tacit assumption that chaos itself depends on
the space-time coordinates. In general relativity, this noninvariance
would imply that chaos is a property of the coordinate system rather
than a property of the physical system (see Ref.~\cite{book:94}, and references
therein).

In this Letter we investigate the transformation laws of Lyapunov
exponents under space-time reparametrizations that preserve conditions ({\it i}-{\it iv}). 
To be specific we consider an Euclidean phase space (not to be confused with the pseudo-Riemannian
spacetime), where the relevant invariant measure is the natural measure.
Our principal result is that
Lyapunov exponents transform according to
\begin{equation}
h^i_{\tau}={ h^i_t }/{ \langle\lambda\rangle_{t} }\; \; \;\;\; (i=1,\ldots ,N),
\label{2}
\end{equation}
where $0<\langle\lambda\rangle_{t}<\infty$ is the time average of
$\lambda=d\tau/dt$ over typical trajectories and $N$ is the
phase-space dimension \cite{dim}. The {\it values} of the Lyapunov exponents are,
of course, noninvariant because a simple reparametrization such as
$(\mathbf{ x},t)\rightarrow (\mathbf{ x}, \alpha t)$ transforms the
exponents $h^i$ into $h^i/\alpha$. However, the {\it signs} of the
Lyapunov exponents are invariant. In particular, if $h>0$ is the
largest Lyapunov exponent, the reparametrization only changes the
characteristic time-scale ${\cal T}\equiv 1/h$ for the manifestation of the
chaotic behavior.  The striking implication of our findings is that
{\it chaos}, as characterized by positive Lyapunov exponents, {\it is
coordinate independent}. As we show, the vanishing of Lyapunov
exponents in the mixmaster cosmology as well as in other examples
previously considered in the literature is due to the violation of at
least one of the conditions ({\it i-iv}) above, which are required for
the Lyapunov exponents to be meaningful indicators of
initial-condition sensitivity and hence chaos.  For instance, it has
been frequently claimed that the exponential divergence of initially
close trajectories that separate as $\delta \mathbf{ x}(t) =\delta
\mathbf{ x}_0 \exp (h t)$ can be removed with a logarithmic
reparametrization of the time \cite{c96_97}. The suggested transformation is
defined as $t=\ln\tau$, so that $\delta \mathbf{ x}(\tau) =\delta
\mathbf{ x}_0 \tau^h$.  This reparametrization is then interpreted as
converting a positive Lyapunov exponent into a zero Lyapunov exponent,
and the inverse of this transformation has been used to support the
claim that integrable systems can have positive Lyapunov exponents \cite{rb01}.
The problem with this argument is that, if the original system is
autonomous, the reparametrized system is necessarily nonautonomous.
Alternatively, if we increase the dimension of the phase space in
order to eliminate the explicit dependence on time, the orbits become
unbounded.  In any case, Lyapunov exponents are not valid indicators
of chaos. We show that a similar problem, although more subtle, is
present in the mixmaster cosmological model.

The Lyapunov exponents of an invariant set of the phase space of
system (\ref{1}) are defined as
\begin{eqnarray}
h^i_t=\lim_{t\rightarrow\infty}\frac{1}{t} \ln\frac{| \boldsymbol{
\eta}^i_t(t)|}{| \boldsymbol{ \eta}^i_{t 0}|}\;\; \;\;\; (i=\!&1,\ldots , N),
\label{3}\\
\mbox{where }\;\;\;\; {d\boldsymbol{ \eta}^i_t(t)}/{dt}=\mathbf{ DF}(\mathbf{
x}(t))\cdot\boldsymbol{ \eta}^i_t(t),
\label{4}
\end{eqnarray}
$\mathbf{ x}(0)$ is a typical initial condition, and
$\boldsymbol{ \eta}^i_{t0}=\boldsymbol{ \eta}^i_{t}(0)$ are tangent vectors
at $\mathbf{ x}(0)$.
We assume that $\mathbf{ F}$ is a continuously differentiable function of $\mathbf{x}$
and system (\ref{1}) has $N$ independent Lyapunov exponents.
Behind definition (\ref{3}-\ref{4}) are the hypotheses ({\it
i-iv}), namely, that with respect to $(\mathbf{ x},t)$, the function
$\mathbf{ F}$ does not depend explicitly on $t$, the dynamics is well
defined for $t$ in the interval $[0,\infty)$, and the invariant set is
bounded and has finite natural measure \cite{measure}.  Otherwise, positive Lyapunov
exponent is not a well defined criterion for chaos. To ensure that
system (\ref{1}) remains autonomous after the coordinate
transformation $(\mathbf{ x},t)\rightarrow (\mathbf{ y},\tau)$, we
consider that when functions $\lambda$ or $\boldsymbol{ \psi}$ depend
explicitly on $t$, the coordinates $\mathbf{x}$ and $\mathbf{y}$ are
redefined to incorporate $t$ as an additional dimension in the phase
space \cite{boundness}.  As a result, the coordinate transformation is
always reduced to a time-independent transformation of the form:
\begin{equation}
\mathbf{ y}=\boldsymbol{ \psi}(\mathbf{ x}),\;\;\; d\tau =\lambda(\mathbf{ x})dt,
\label{5}
\end{equation} 
where $\lambda$ is a strictly positive, continuously differentiable function,
and $\boldsymbol{ \psi}$ is a diffeomorphism.  This is the general
class of transformations for which integrability is coordinate
invariant \cite{ml02} in the sense that, if $\{I_1,I_2,\ldots \}$ are independent
integrals of motion with respect to the coordinates $(\mathbf{ x},t)$,
then $\{I_1\circ\boldsymbol{
\psi}^{-1}, I_2\circ\boldsymbol{ \psi}^{-1},\ldots \}$ are independent
integrals of motion with respect to $(\mathbf{ y},\tau)$.

Transformation (\ref{5}) is composed of a time
reparametrization followed by a space diffeomorphism. It is well
known that the Lyapunov exponents are invariant under space
diffeomorphisms \cite{o94}.  Without loss of generality, we consider only
transformations of the time parameter: $(\mathbf{ x},t)\rightarrow
(\mathbf{ x},\tau)$, where $d\tau =\lambda(\mathbf{ x})dt$.  All the
orbits of the phase space are invariant under this kind of
transformation as the velocity field of the reparametrized flow is
parallel to the original one:
 ${d\mathbf{ x}}/{d\tau}=\lambda^{-1}(\mathbf{ x}) \mathbf{F}(\mathbf{ x})$.
The Lyapunov exponents, however, may be different because
Eqs.~(\ref{3}-\ref{4}) become
\begin{eqnarray}
h^i_{\tau}&=&\lim_{\tau\rightarrow\infty}\frac{1}{\tau} \ln\frac{|
\boldsymbol{ \eta}^i_{\tau}(\tau)|}{| \boldsymbol{ \eta}^i_{\tau 0}|},
\label{8}\\
\mbox{and }\;\;\;\;
{d\boldsymbol{ \eta}^i_{\tau}(\tau)}/{d\tau}&=&\mathbf{
D}[\lambda^{-1}\mathbf{ F}](\mathbf{ x}(\tau))\cdot\boldsymbol{
\eta}^i_{\tau}(\tau),
\label{9}
\end{eqnarray}
respectively. Incidentally, {\it mixing}, which is a property most
often observed in chaotic systems, is {\it not} invariant under
transformation (\ref{5}) since it has been shown that an adequate time
reparametrization of a nonchaotic irrational flow on a 3-torus {\it is}
mixing \cite{f02}.

For the same initial conditions, the defining relations of
$h^i_{\tau}$ and $h^i_t$ present two different factors. The first,
associated with the time average of $\lambda$, comes from
the difference between the two time parametrizations along the same orbit
and is factored out when Eq.~(\ref{8}) is written as
\begin{equation}
h^i_{\tau}=\frac{1}{ \langle\lambda\rangle_{t}}
\lim_{t\rightarrow\infty}\frac{1}{t} \ln\frac{| \boldsymbol{
\eta}^i_{\tau}(t)|} {| \boldsymbol{ \eta}^i_{\tau 0}|},
\label{10}
\end{equation} 
where $\boldsymbol{ \eta}^i_{\tau}(t)\equiv \boldsymbol{ \eta}^i_{\tau}(\tau(t))$,
$\tau(t)\equiv\int_0^t\lambda(\mathbf{ x}(t))dt$, and
$\langle\lambda\rangle_{t}\equiv
\lim_{t\rightarrow\infty}\frac{1}{t}\int_0^{t}\lambda(\mathbf{ x}(t))dt$.
Condition $0<\langle\lambda\rangle_{t}<\infty$ is a basic
requirement for the natural measure to be well defined \cite{measure}.  The second factor,
due to the gradient of $\lambda$, is associated with the difference
between the surfaces of simultaneous time for each
parametrization and is separated out when $\boldsymbol{
\eta}^i_{\tau}$ in Eq.~(\ref{9}) is parametrized in terms of $t$
rather than $\tau$:
\begin{equation}
\frac{d\boldsymbol{ \eta}^i_{\tau}(t)}{dt}= \mathbf{ DF}(\mathbf{
x}(t))\cdot\boldsymbol{ \eta}^i_{\tau}(t)- [\mathbf{
F}\cdot\nabla^{\scriptscriptstyle \dag}\ln\lambda ](\mathbf{ x}(t))\cdot\boldsymbol{
\eta}^i_{\tau}(t).
\label{12}
\end{equation}
The last term in this equation implies that the time evolution of
vector $\boldsymbol{ \eta}^i_{\tau}(t)$ in Eq.~(\ref{10}) is in general
different from that of vector $\boldsymbol{ \eta}^i_{t}(t)$ in
Eq.~(\ref{3}).  But the relevant
question is: how large is this difference?

Here we show that the difference is sub-exponential, in the  sense that
$\boldsymbol{ \eta}^i_{t}(t)$ and  $\boldsymbol{ \eta}^i_{\tau}(t)$
grow or shrink with the same exponential rate. This implies
\begin{equation}
\lim_{t\rightarrow\infty}\frac{1}{t} \ln\frac{| \boldsymbol{\eta}^i_{\tau}(t)|} {| \boldsymbol{ \eta}^i_{\tau 0}|}=
\lim_{t\rightarrow\infty}\frac{1}{t} \ln\frac{| \boldsymbol{\eta}^i_{t}(t)|} {| \boldsymbol{ \eta}^i_{t 0}|},
\label{13}
\end{equation}
which in turn implies our main result (\ref{2}).

First we analyze periodic orbits, which form the most fundamental
building blocks of chaotic sets \cite{periodic_orbits}. On a periodic orbit $\mathbf{
x}^*$, it is convenient to adopt the explicit notation
$\boldsymbol{\eta}^i_{t}(\mathbf{x}^*(t),t)$ and
$\boldsymbol{\eta}^i_{\tau}(\mathbf{x}^*(t),t)$ for the solutions of the
variational Eqs.~(\ref{4}) and (\ref{12}), respectively.  If
$\mathbf{x}^*$ is a fixed point, we trivially have
$\boldsymbol{\eta}^i_{\tau}(\mathbf{x}^*(t),t)=\boldsymbol{\eta}^i_{t}(\mathbf{x}^*(t),t)$
because the last term in Eq.~(\ref{12}) is zero. Now consider that $\mathbf{x}^*$ is a periodic orbit with
least period $T>0$ with respect to $t$, so that
$\mathbf{x}^*(t+T)=\mathbf{x}^*(t)$.  Let $h_t^i(\mathbf{x}^*)$ denote
the {\it local} Lyapunov exponents for Eq.~(\ref{3}) on
$\mathbf{x}^*$, and $\sigma_t^i(\mathbf{x}^*) \equiv
\exp(h_t^i(\mathbf{x}^*))$ denote the corresponding local Lyapunov numbers,
where $i=1,\ldots ,N$.  One of the Lyapunov numbers, say $\sigma_t^N$, is 1
because the bounded function $d\mathbf{x}^*/dt$ is a solution of
Eq.~(\ref{4}), rendering zero to the corresponding Lyapunov exponent.
The same is true for any parametrization.  To study the other
Lyapunov numbers, let $\pi$ be the hyperplane orthogonal to $\mathbf{
F}(\mathbf{ x}^*(0))$ at $\mathbf{ x}^*(0)$, and
$\mathbf{M}: U\subset\pi\mapsto\pi$ be the first return map on this hyperplane,
defined in a neighborhood $U$ of $\mathbf{ x}^*(0)$.  This map does
not depend on the time parametrization of the continuous flow, being
exactly the same for both the original and
reparametrized flow.  It follows from standard results in
Floquet theory \cite{floquet} that the local Lyapunov numbers
$\sigma_M^i(\mathbf{x}^*(0))$ of this map, defined as the
 magnitude of the eigenvalues of the Jacobian matrix of $\mathbf{M}$ at $\mathbf{x}^*(0)$, are
the power $T$ of the first $N-1$ local Lyapunov numbers of the flow, i.e.,
$\sigma_M^i(\mathbf{x}^*(0))= \sigma_t^i(\mathbf{x}^*)^T$ for
$i=1,\ldots ,N-1$.  Geometrically, the local Lyapunov numbers of $\mathbf{M}$
can be defined as $\sigma_M^i(\mathbf{x}^*(0))=\exp(h_M^i(\mathbf{x}^*)
T)$, where $h_M^i(\mathbf{x}^*)$ is defined through Eq.~(\ref{3}) with
$\boldsymbol{\eta}^i_{t}(\mathbf{x}^*(t),t)$ replaced with its orthogonal
component to
$\mathbf{F}(\mathbf{x}^*(t))$.  From the identity $h_M^i(\mathbf{x}^*)=h_t^i(\mathbf{x}^*)$
then follows $ h_M^i(\mathbf{x}^*)=
\lim_{t\rightarrow\infty}\frac{1}{t}\ln\frac{|\boldsymbol{\eta}^i_{t}(\mathbf{x}^*(t),t)|}{|\boldsymbol{
\eta}^i_{t}(\mathbf{x}^*(0),0)|} $.  The same is true for
$\boldsymbol{\eta}^i_{\tau}(\mathbf{x}^*(t),t)$, so that if
$\tilde{h}_t^i$ denotes the limit in Eq.~(\ref{10}) then
$h_M^i(\mathbf{x}^*)=\tilde{h}_t^i(\mathbf{x}^*)$ and
$ h_M^i(\mathbf{x}^*)=
\lim_{t\rightarrow\infty}\frac{1}{t}\ln\frac{|\boldsymbol{\eta}^i_{\tau}(\mathbf{x}^*(t),t)|}
{|\boldsymbol{ \eta}^i_{\tau}(\mathbf{x}^*(0),0)|}.
$ 
From these, it follows that Eq.~(\ref{13}) holds on periodic orbits.

But the same must be valid in general because
the Lyapunov exponents of typical orbits are weighted averages over the local Lyapunov
exponents of all periodic orbits in the respective ergodic component.
The weight is the fraction of time
spent by a typical orbit near the corresponding periodic orbit and is
uniquely determined by the largest local Lyapunov exponent (assumed to
be positive). The smaller this Lyapunov exponent the longer it takes
for the orbit to move away from that neighborhood. In other words, the
largest local Lyapunov exponents define a natural measure, over which
all the Lyapunov exponents are computed.  Since
$\tilde{h}_t^i(\mathbf{x}^*)=h_t^i(\mathbf{x}^*)$, both the measure
and the local exponents are the same for $h_t^i$ and $\tilde{h}_t^i$.
Therefore $\tilde{h}_t^i=h_t^i$ on typical orbits, and this is
equivalent to Eq.~(\ref{13}).

We now show that similar arguments can be extended directly to typical
orbits.  Eq.~(\ref{4}) can be written as a map $M(\boldsymbol{
\eta}^i_t(t))=\boldsymbol{ \eta}^i_t(t+\delta t)$, where $\boldsymbol{
\eta}^i_t(t+\delta t)= \boldsymbol{ \eta}^i_t(t)+\delta t\mathbf{
DF}(\mathbf{ x}(t))\cdot\boldsymbol{ \eta}^i_t(t)$.  For the natural
measure to be normalizable, $\mathbf{ F}(\mathbf{ x}(t))$ and
$\nabla|\mathbf{F} (\mathbf{ x}(t))|$ must not
grow exponentially and $\mathbf{ F}(\mathbf{ x}(t))$ must not shrink
exponentially along typical orbits.  Since $d\mathbf{ x}/dt$ is a
solution of Eq.~(\ref{4}), the component $P(\boldsymbol{ \eta}^i_t)$ of
$\boldsymbol{ \eta}^i_t$ parallel to the flow remains parallel in each
iteration, and the Lyapunov exponent in this direction is zero.  The
orthogonal component $Q(\boldsymbol{ \eta}^i_t)$ of $\boldsymbol{
\eta}^i_t$ is mapped into two parts, one parallel and the other
orthogonal to the flow. The parallel part is 
$PMQ(\boldsymbol{ \eta}^i_t(t))=
\delta t \boldsymbol{[}\mathbf{F}\cdot\nabla^{\scriptscriptstyle \dag}\ln|\mathbf{F}|\boldsymbol{]} (\mathbf{ x}(t))\cdot Q(\boldsymbol{ \eta}^i_t(t))$,
where neither 
$\mathbf{F}(\mathbf{ x}(t))$ nor $\nabla\ln|\mathbf{F} (\mathbf{ x}(t))|$ grows exponentially with $t$.
If $h_t^i<0$ ($h_t^i>0$), each of
the projections $P(\boldsymbol{ \eta}^i_t)$ and $Q(\boldsymbol{
\eta}^i_t)$ must shrink at least (grow at most) as $\exp(h_t^it)$.  But
$Q(\boldsymbol{ \eta}^i_t)$ cannot shrink faster (grow slower) than $P(\boldsymbol{
\eta}^i_t)$ because otherwise the contribution of the orthogonal part
to Eq.~(\ref{3}) would be negligible, resulting in $h_t^i=0$, which
violates the hypothesis that $h_t^i\neq 0$. Therefore,
$Q(\boldsymbol{ \eta}^i_t)\sim \exp(h_t^it)$ for both $h_t^i<0$ and $h_t^i>0$. 
A similar relation is valid for $\boldsymbol{
\eta}^i_{\tau}(t)$ because $\lambda^{-1}d\mathbf{ x}/dt$ is a solution
of Eq.~(\ref{12}).  We then compare the solutions of Eq.~(\ref{12})
with those of Eq.~(\ref{4}) for identical initial conditions.  The
term involving matrix $-[\mathbf{ F}\cdot\nabla^{\scriptscriptstyle \dag}\ln\lambda](\mathbf{
x}(t))$ is parallel to the flow and does not affect the orthogonal
part in Eq.~(\ref{12}).  The term that contributes to the orthogonal
component, $\mathbf{ DF}(\mathbf{ x}(t))$, is the same as that
in Eq.~(\ref{4}).  Therefore, $Q(\boldsymbol{
\eta}^i_t(t))=Q(\boldsymbol{ \eta}^i_{\tau}(t))$, which again leads to
Eq.~(\ref{13}).

An important implication of our findings is that positive Lyapunov
exponents are necessarily mapped into positive Lyapunov exponents
under time reparametrizations. This implies that the previous
examples of noninvariant chaos in cosmology are based on the violation
of hypotheses required for an interpretable computation of the
Lyapunov exponents. Consider, for example, the mixmaster cosmological
model \cite{m69}, which is believed to describe generic cosmological singularities,
and  whose relevant hypotheses can be discussed explicitly.
In the asymptotic limit (close to the big bang),
the essential features of the continuous dynamics are represented in the Farey map,
$F(u)=u-1$ if $u\geq 1$ and $F(u)=u^{-1}-1$ if $u<1$,
whose Lyapunov exponent is zero. This result is claimed to be in
conflict with the corresponding result for the first return map on $[0,1]$
(Gauss map),
$G(v)=1/v-[1/v]$, where $[1/v]$ is the integer part of $1/v$,
whose Lyapunov exponent is positive \cite{b82}: $h=\pi^2/6\ln 2$.
The problem here is that, different from the first return map,
the orbits of map $F$ are typically unbounded.
Map $F$ can be compactified for $w=(u+1)^{-1}$,
by defining $H:[0,1]\rightarrow [0,1]$,
$H(w)=w/(1-w)$ if $w\leq 1/2$ and $H(w)=(1-w)/w$  if $w>1/2$.
But the Lyapunov exponent is still zero. The problem now is that the
invariant density, $\rho (w)=1/w$, is not normalizable and all the
contribution to the Lyapunov exponent comes from points that are
arbitrarily close to $w=0$. Similar problems are present in the
continuous dynamics since in the usual coordinates the model is
either nonautonomous  or noncompact \cite{m69}. In addition, because the dynamics is
limited by a cosmological singularity, the domain of frequently used
time parameters, such as the cosmological time and the volume of the
universe, is necessarily finite.

From the above it appears that autonomous equations, bounded motions, and
normalizable measure are properties mutually incompatible in the
mixmaster dynamics when all the orbits are taken into account.
We observe, however, that there are invariant
bounded subsets of orbits in map $F$ as well as invariant subsets with
normalizable measure in map $H$ which do have positive
Lyapunov exponents. For example, map $H$ has a nontrivial
set of invariant orbits embedded in the
interval $[\alpha,(1+\alpha)^{-1}]$, for every $\alpha\in (0,1/3]$,
which has normalizable measure. The invariant set is composed
of all the orbits that never leave this interval and as such
contains a countable number of periodic orbits and an uncountable
number of nonperiodic orbits. The
Lyapunov exponent of the invariant set, as computed along typical nonperiodic orbits,
satisfies $h \ge 2\ln(1+\alpha)$.
Since map $H$ corresponds to the asymptotic
behavior of one parametrization of the continuous mixmaster model, an
invariant set of map $H$ must correspond to an invariant
set of the asymptotic dynamics for this particular parametrization,
which therefore satisfies conditions ({\it i}-{\it iv}) and has positive Lyapunov
exponent.
It follows then from our main result (\ref{2}) that these asymptotically invariant
sets must have positive Lyapunov exponents for any space-time reparametrization of
the continuous dynamics that preserves these conditions on them.
These invariant sets are therefore chaotic with respect to any coordinate system
for which the Lyapunov exponents can be properly computed.
In these sense one can meaningfully say that the mixmaster cosmology exhibits {\it
coordinate independent chaotic behavior} close to the big bang.

Summarizing, we have uncovered the transformation laws of the Lyapunov
exponents for flows under space-time reparametrizations. Strikingly,
systems exhibiting exponential separation of nearby orbits with respect to
one choice of the time parameter will display exponential divergence
with respect to any other time parameter that preserves conditions
({\it i}-{\it iv}). This implies that chaos is invariant under time
reparametrizations, which is in sharp contrast with previous results
in relativistic cosmology, where the apparent noninvariance of chaos
has been the subject of an intensive debate. Our
findings thus shed new light on the {\it conceptual} problem of chaos
in cosmology \cite{saddle}.


\begin{references}
\bibitem{dim}
In this Letter we focus on general relativistic systems with
finite number of degrees of freedom. However, since the
essential dynamics of a PDE may be captured by finite expansions,
our framework is also potentially relevant for models with
infinite degrees of freedom. The number of degrees of
freedom does not appear to be an issue for the possible
noninvariance of chaos in general relativity.
\bibitem{measure}
We exclude the cases where the measure accumulates on trivial subsets of the invariant set,
         such as isolated points and simple boundaries \cite{floquet}.
\bibitem{boundness} 
The orbits in the extended phase space will be
unbounded along $t$. But when the transformed system depends only
periodically on $t$, this can be remedied by taking $t$~mod~$p$
rather than $t$ as the new phase space coordinate, where $p$ is
the period of the time dependence. Lyapunov exponents can also be
defined for some nonautonomous systems
[J.-P. Eckmann and D. Ruelle, Rev. Mod. Phys. {\bf 57}, 617 (1985)].
\end{references}
\end{document}